\documentclass[]{aa}  

\usepackage{natbib}
\bibpunct{(}{)}{;}{a}{}{,}
\usepackage{graphicx}
\usepackage{revsymb}	
\usepackage{amsmath}	
\usepackage[usenames]{color}	
\usepackage{epstopdf}	
\usepackage{txfonts}
\usepackage{amssymb}
\usepackage{color}

\newcommand{\ind}[1]{_{\mathrm{#1}}}
\newcommand{\numax}{\nu\ind{max}}
\newcommand{\Tobs}{T\ind{obs}}

\begin{document}

   \title{Theoretical power spectra of mixed modes \\in low mass red giant stars}
   \author{M. Grosjean\inst{1}  \and M.-A. Dupret\inst{1} \and K. Belkacem \inst{2} \and J. Montalban\inst{1} \and R. Samadi\inst{2} \and B. Mosser \inst{2}}
  \institute{Institut d'Astrophysique et de G\'eophysique, Universit\'e de Li\`ege, All\'ee du 6 Ao\^{u}t 17, 4000 Li\`{e}ge, Belgium  \\ \email{grosjean@astro.ulg.ac.be}
\and LESIA, Observatoire de Paris, CNRS UMR 8109, Universit\'e Paris Diderot, 5 place J. Janssen, 92195 Meudon, France }

 \abstract
   {CoRoT and \emph{Kepler} observations of red giant stars revealed very rich spectra of non-radial solar-like oscillations. Of particular interest was the detection of mixed modes that exhibit significant amplitude, both in the core and at the surface of the stars. It opens the possibility of probing the internal structure from their inner-most layers up to their surface along their evolution on the red giant branch as well as on the red-clump. 
}
   {Our objective is primarily to provide physical insight into the physical mechanism responsible for mixed-modes amplitudes and lifetimes. Subsequently, we aim at understanding the evolution and structure of red giants spectra along with their evolution. The study of energetic aspects of these oscillations is also of great importance to predict the mode parameters in the power spectrum. 
}
   { 
   Non-adiabatic computations, including a time-dependent treatment of convection, are performed and provide the lifetimes of radial and non-radial mixed modes. We then combine these mode lifetimes and inertias with a stochastic excitation model that gives us their heights in the power spectra.}
{For stars representative of CoRoT and \emph{Kepler} observations, we show under which circumstances mixed modes have heights comparable to radial ones. We stress the importance of the radiative damping in the determination of the height of mixed modes. Finally, we derive an estimate for the height ratio between a g-type and a p-type mode. This can thus be used as a first estimate of the detectability of mixed-modes. }
   {}

   \maketitle
%

\section{Introduction}

One of the major achievements of the CoRoT \citep{Baglin2006} and \emph{Kepler} \citep{Gilliland2010} space-borne missions has been the detection of a rich harvest of both radial and non-radial solar-like oscillations in red-giant stars \citep[e.g.][]{DeRidder09,Mosser10,Bedding11,Beck11}. These stars present a structure characterized by a high density contrast between the core and the envelope. It leads to the appearance of modes behaving as acoustic modes in the stellar envelope
and as gravity modes in the core.These \emph{mixed modes}, as named in the early works of  \citet{Dziembowski71} for Cepheids and \citet{Scuflaire74} for condensed polytropic model, have been subject to an  extensive investigation from a theoretical point of view \citep[e.g.][]{Dziembowski01,Dupret09,Montalban10,Dziembowski12,Montalban13}. From an observational point of view, these modes present the major advantage of having detectable amplitudes at the star surface and of being able to probe the inner-most region.  We called the modes that present acoustic modes characteristics p-type modes. The others, which are mostly trapped in the core are called g-type modes.

Among other results, the period spacing of mixed-modes enables to determine the stars' evolutionary stage \citep[][]{Bedding11,Mosser11}. Indeed, it allows to distinguish between the stars belonging to the ascending red-giant branch (hydrogen shell-burning phase) from those belonging to the red-clump (helium central burning phase), which was previously impossible for field stars. This leap forward then opened the way to a large number of applications. For instance, it permitted to give constraints on the mass loss during the helium flash \citep{Mosser12}, to investigate stellar population in the Galaxy \citep[e.g.][]{Miglio09,Miglio13}, and more recently to identify the physical nature of the semi-regular variability in M giants \citep{Mosser13b}. Another key result is the ability of mixed modes to unveil the rotation profile of the inner-most region of red giants \citep[][]{Beck2012,Deheuvels2012,Mosser2012c,Goupil13,Ouazzani2013,Deheuvels2014} as well as to emphasize the need for additional physical processes to transport angular momentum in  current red giants models \citep[e.g.][]{Eggenberger2012,Marques2013}. 

All these works were mainly focused on the red giant oscillation pattern. However, the observed power spectra also provides us with additional information through mode heights and linewidths which are related to the amplitudes and lifetimes of the modes (see Sect. \ref{sect:nonad} and \ref{sect:stoc} for details).
Those observables have recently been considered more thoroughly for radial modes both on the observational side \citep[e.g.][]{Baudin2011,Corsaro2012,Appourchaux2012,Corsaro2013,Appourchaux2014} and on the theoretical side \citep[e.g.][]{Chaplin2009,Belkacem11,Belkacem12} since they are important for estimates of mode detectability and give us constraints on the interaction between pulsation and turbulent convection. 
Nevertheless, the literature is more tenuous about the amplitudes and lifetimes of mixed-modes. Only very recent measurements of mixed-modes linewidths and amplitudes have been performed by \cite{Benomar2013,Benomar2014}. This is mainly due to the complexity of mode fitting for power spectra including mixed modes. Moreover, heights and linewidths determined from observed spectra are closely correlated \citep[see][]{Chaplin98}. Concerning theoretical modelling of the these modes, \cite{JCD04} discussed the trapping and inertia of non-radial adiabatic modes and the possible effects on amplitudes of subgiants and red giants.\cite{HG02} computed theoretical amplitudes and lifetimes of radial modes in the star $\xi$ Hydrae. Finally, \cite{Dupret09}, based on non-adiabatic calculations as well as modelling of mode excitation, proposed the first theoretical modeling of the observed power spectra including dipole and quadrupole mixed modes. It enables us to gain insight into the interpretation of the observed spectra of red giants and their change with stellar evolution. However, this work was restricted to quite massive models ($2$ and $3 M_\odot$). 

Motivated by all these recent results and the pioneering work of \cite{Dupret09}, our objective is to go further in the theoretical study of the energetic aspects of these oscillations in red giants.
We assume that the modes are stochastically excited by turbulent motions at the top of the convective envelope \citep[see e.g.][]{Samadi11} and damped through the coherent interaction between convection and oscillations \citep[see e.g.][]{BelkacemSamadi13}.
In this paper, we present models of lower masses than in \citet{Dupret09}, more representative of CoRot and \emph{Kepler} samples, and discuss the effect of the mass on the detectability of mixed modes. Note that compared to \citet{Dupret09}, both our equilibrium and oscillations models  have been highly improved for the specific case of red giants (see Sect.~\ref{sect:mod_method}).
The paper is thus organized as follows: in Sect.~\ref{sect:mod_method}, we compute lifetimes of radial and non-radial modes with a non-adiabatic code, using a non-local and time dependent treatment of convection  as well as the amplitudes with a stochastic excitation model. With these results, we discuss in Sect.~\ref{sect:results} the effect of the radiative damping on the height ratio between a g-type and p-type mixed mode. We also study the effect of the duration of observation on our synthetic power spectra to draw conclusions about the possibility to detect mixed modes.
Finally, Sect.~\ref{sect:discussion} is dedicated to conclusions and discussions.

\section{Modelling of the power spectra} 
\label{sect:mod_method} 
\subsection{Computation of equilibrium models}
\label{sect:models}

\begin{figure}[h!]
   \centering
     \includegraphics[width=1.\linewidth]{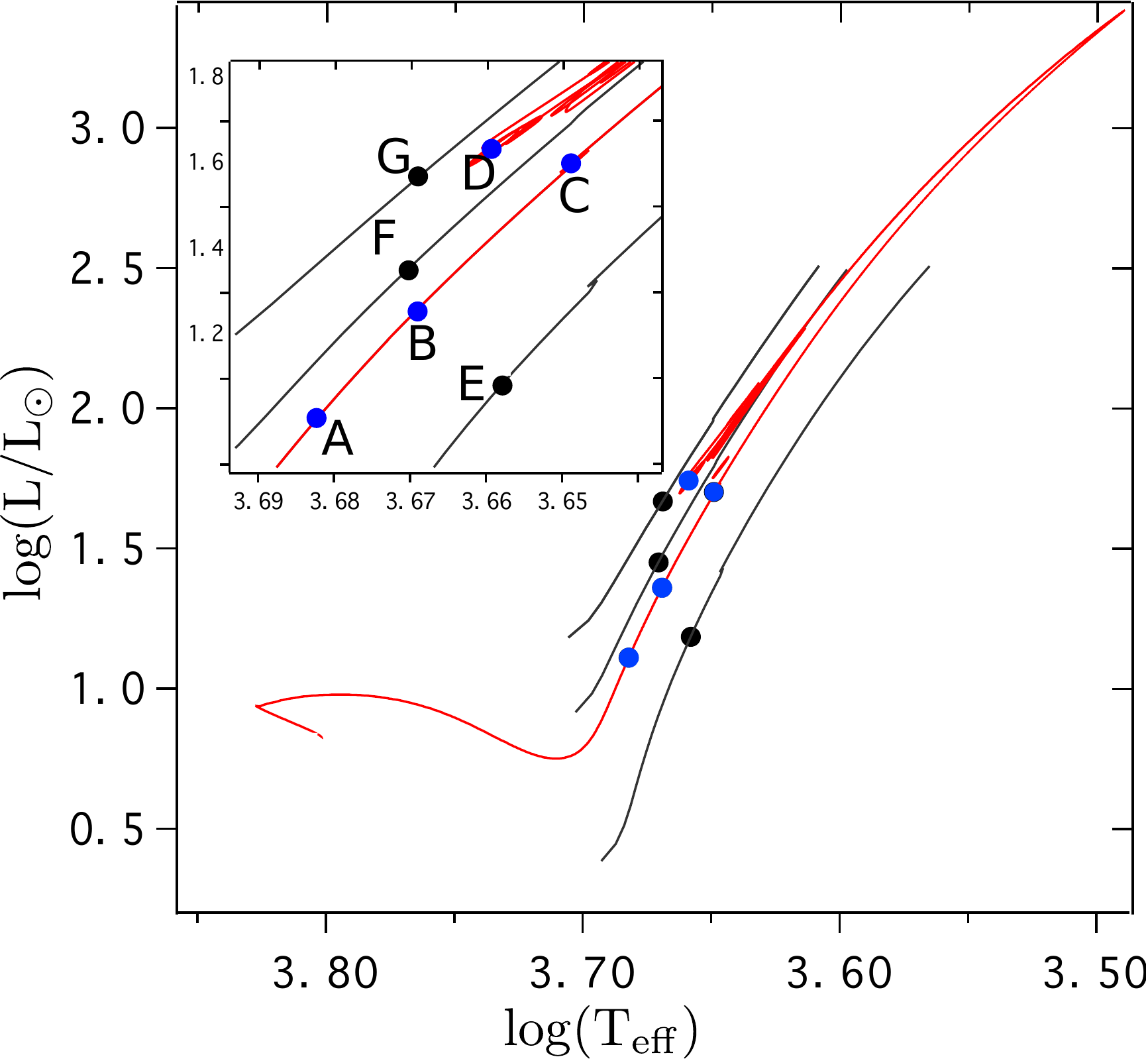}
      \caption{Evolutionnary tracks in the Hertzprung-Russel diagram of our models. Selected models are represented by dots. Blue dots corresponds to models of a $1.5 M_\odot$ star at different ages (on the red-giant branch and in the clump). Black dots corresponds to models with the same number of mixed-modes over a large separation.
              }
         \label{Fig:HR}
   \end{figure}

We first consider $1.5 M_\odot$ models which are typical of CoRoT and \emph{Kepler} observed red-giant stars, from the bottom of the red-giant branch to the helium core-burning phase (see models A to D in Table.~\ref{tab:models} and Fig.~\ref{Fig:HR}). 
We give for each models the global seismic parameters, i.e. the large frequency separation ($\Delta\nu$), the frequency of maximum oscillation power ($\numax$) and the asymptotic period spacing ($\Delta \Pi$).
Note that an adiabatic analysis of these models is presented in \citet{Montalban13}. 
Secondly, we selected models between 1 and 2.1 $M_\odot$  (see models E to G in Table.~\ref{tab:models} and Fig.~\ref{Fig:HR}) at a similar evolutionary stage as model B. The criteria to choose these models as well as the consquences on theoretical power spectra are discussed in Sect. \ref{sect:discussion}.

All the equilibrium models have been computed using the ATON stellar evolutionary code \citep{Ventura08} with $X=0.7$ and $Z=0.02$ for the initial chemical composition. The convection is described by the classical mixing-length theory \citep{BohmVitense58} with $\alpha_{MLT}=1.9$. 
The radiative opacities come from OPAL \citep{Iglesias96} for the metal mixture of \citet{GN93} completed with \citet{Alexander94} at low temperatures. The conductive opacities correspond to the \citet{Potekhin99} treatment corrected following the improvement of the treatment of the e-e scattering contribution \citep{Cassisi07}.
Thermodynamics quantities are derived from OPAL \citep{Rogers02}, \citet{Saumon95} for the pressure ionisation regime and \citet{Stolzmann96} treatment for the He/C/O mixtures. 
Finally, the nuclear cross-sections are from NACRE compilation \citep{Angulo99}, and the surface boundary conditions are provided by a grey atmosphere following the treatment by \citet{Henyey65}.

\begin{table}[!h]
\caption{Global parameters of our models }
  \begin{center}
\scalebox{0.83}
 {
  \begin{tabular}{cccccc}\hline 
{Model} & {Mass [$M_\odot$]} & {Radius  [$R_\odot$] } & {$\Delta \Pi$ [s] } & {$\Delta \nu \left[\mu Hz\right] $}  &{$\numax \left[ \mu Hz\right]$ }\\  \hline
   A & 1.5 & 5.17 &79.7  &  14.1&190 \\
   B & 1.5& 7.31 & 70.5 &  8.4 & 97 \\
   C & 1.5 & 11.9 & 57.2  & 4 &37 \\  \hline
   D & 1.5 & 11.9 & 242.5 &  4 & 37\\ \hline
   E & 1.0 & 6.3 &  76.7& 8.5 & 88 \\ 
   F & 1.7 & 8.1 & 68.6 &  7.7&90\\ 
   G &  2.1 & 10.5 & 88.0 &  5.7&66 \\ \hline
  \end{tabular}
  }
  \label{tab:models}
\tablefoot{
 The large separation $\Delta\nu$ and the frequency of maximum power $\numax$ are computed using the seismic scaling relations \citep[e.g.][]{Mosser10,Belkacem12b}. The period spacing $\Delta\Pi$ is computed using the asymptotic expansion \citep[e.g.][]{Tassoul80}. 
}
 \end{center}
\end{table}

\subsection{Non-adiabatic computations}
\label{sect:nonad}
To compute theoretical mode frequencies ($\nu$), mode inertias ($I$) and mode lifetimes ($\tau$), we use the non-adiabatic pulsation code MAD \citep{Dupret02}.
The major outcome of non-adiabatic computations being the mode lifetimes (or equivalently damping rates $\eta$), we will describe the different contributions to the damping and the way they are modelled.
The lifetime of a mode, $\tau$, is related to the linewidth $\Gamma$ of the peak in the power spectrum by
\begin{equation}
\label{tau-Gamma}
\Gamma = 1/\pi\tau = \eta/\pi
\end{equation} 

The damping rate of a mode is given by the integral expression  \citep[e.g.][]{Dupret09}
\begin{equation}
\eta=- \frac{1}{2\nu I | \xi_r (R) |^2 M} \; \int_V {\rm d} W \, , 
\label{damping}
\end{equation}
where $\int_V {\rm d} W$ is the work performed by the gas during one oscillation cycle, $\xi_r$ the radial component of the eigendisplacement vector, $R$ and $M$ the total radius and mass of the star, and $I$ the dimensionless mode inertia.

\begin{figure}[t]
    \includegraphics[width=1.\linewidth]{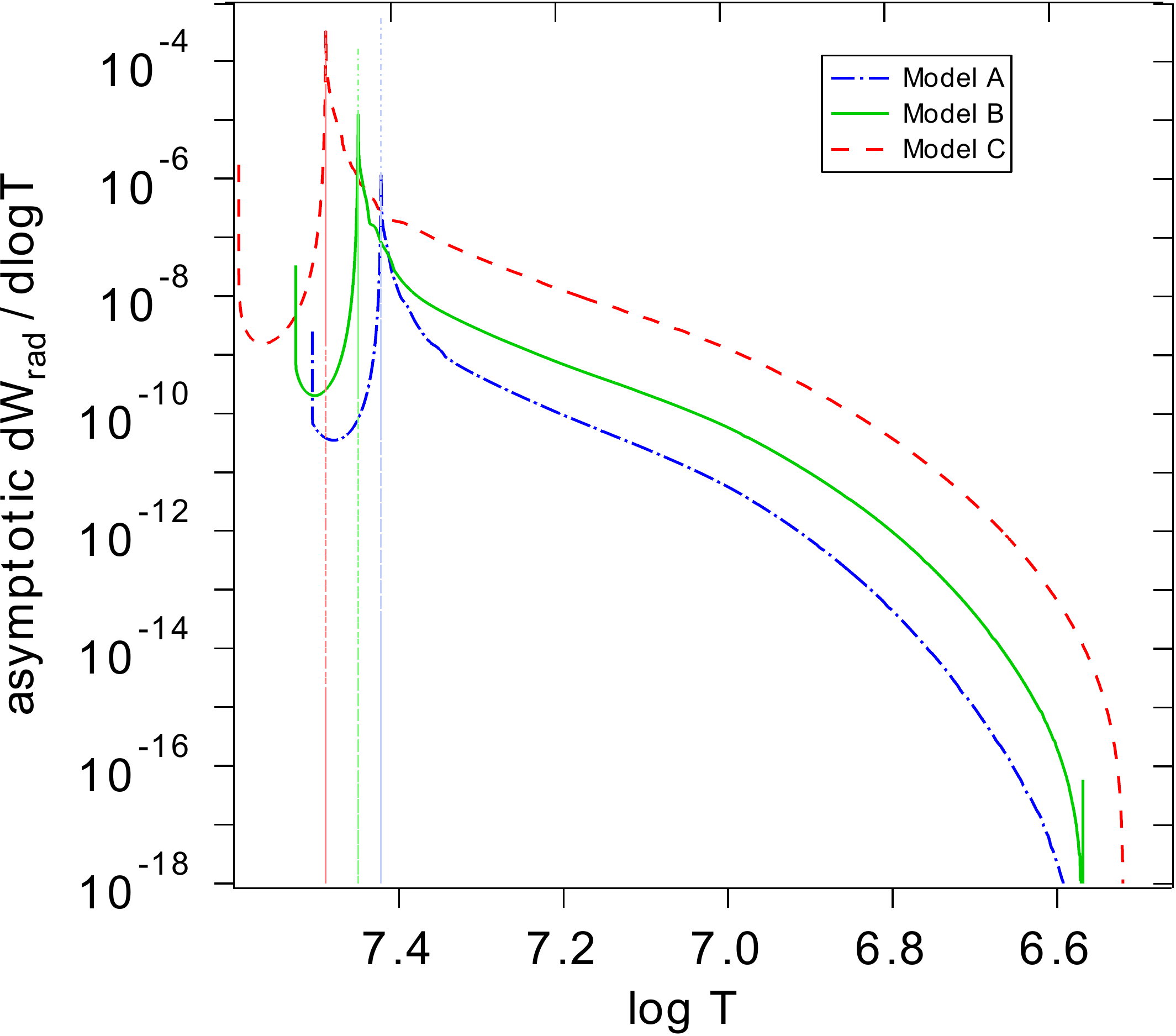} 
      \caption{ Integrant of the asymptotic expansion of the radiative damping (eq. \ref{raddamp}) for models A (blue), B (green) and C (red). The vertical lines represent the lower limit of the H-burning shell in each model. The work is normalised by $GM^2/R$}
         \label{fig:raddamp}
  \end{figure}

Two regions of the red giants can play a significant role in the work integral: the radiative region in the core of the red giant and especially around the bottom of the H-burning shell ($W_c$) and the outer non-adiabatic part of the convective envelope ($W_e$)
\begin{equation}
\int_V {\rm d}W = W _c+ W_e \, .
\label{eq:work}
\end{equation}
To get more insight into the radiative term, it is useful to consider the asymptotic formulation developed by \citet{Dziembowski77} \citep[see also][]{VanHoolst98,Godart09} for $g$ modes. It gives 
\begin{equation}
- \int_{r_0}^{r_{\rm c}} \frac{{\rm d}W}{{\rm d}r}{\rm d}r \simeq \frac{K(\ell (\ell+1))^{3/2}}{2\nu^3} \int_{r_0}^{r_{\rm c}}\frac{\nabla_{\rm ad}-\nabla}{\nabla} \frac{\nabla_{\rm ad} N g L}{p r^5} \, {\rm d}r \, , 
\label{raddamp}
\end{equation}
where $r_0$ and $r_c$ are the lower and upper radius of the g-cavity, $K$ is a normalisation constant, $\nabla$ and $ \nabla_{ad}$ are the real and adiabatic gradients, $N$ the Brunt-V\"{a}is\"{a}l\"{a} frequency, $g$ the gravity, $L$ the local luminosity and $p$ the pressure.
This formulation shows that the main contribution to the radiative damping occurs around the bottom of the H-burning shell. When the star evolves on the red giant branch, $N/r^5$ increases, as a result of the contraction of the central layers, leading to an increase of the radiative damping as shown in Fig.~\ref{fig:raddamp}.

About the convective contribution, we note that the transition region occurs in the upper part of the convective envelope. In this region, the time-scale of most energetic turbulent eddies is also of the same order as the oscillation periods. Hence, it is important for the estimate of the damping rates to take the interaction between convection and oscillations into account. This is made by using a non-local, time-dependent treatment of the convection (TDC) that takes into account the variations of the convective flux and of the turbulent pressure due to the oscillations \citep[see][for the description of this treatment]{Grigahcene05,Dupret06a}. 
\cite{Gough77} proposed a second treatment based on the "kinetic of gas" picture of the MLT.
For the non-local parameters, we used  $a= 10$ and  $b=3 $ according to the definition of \citet{Balmforth92}. This set of parameters fits the turbulent pressure in sub-adiabatic atmospheric layers of a solar hydrodynamic simulation \citep{Dupret06b}.

\begin{figure}[t]
   \centering
     \includegraphics[width=0.9\linewidth]{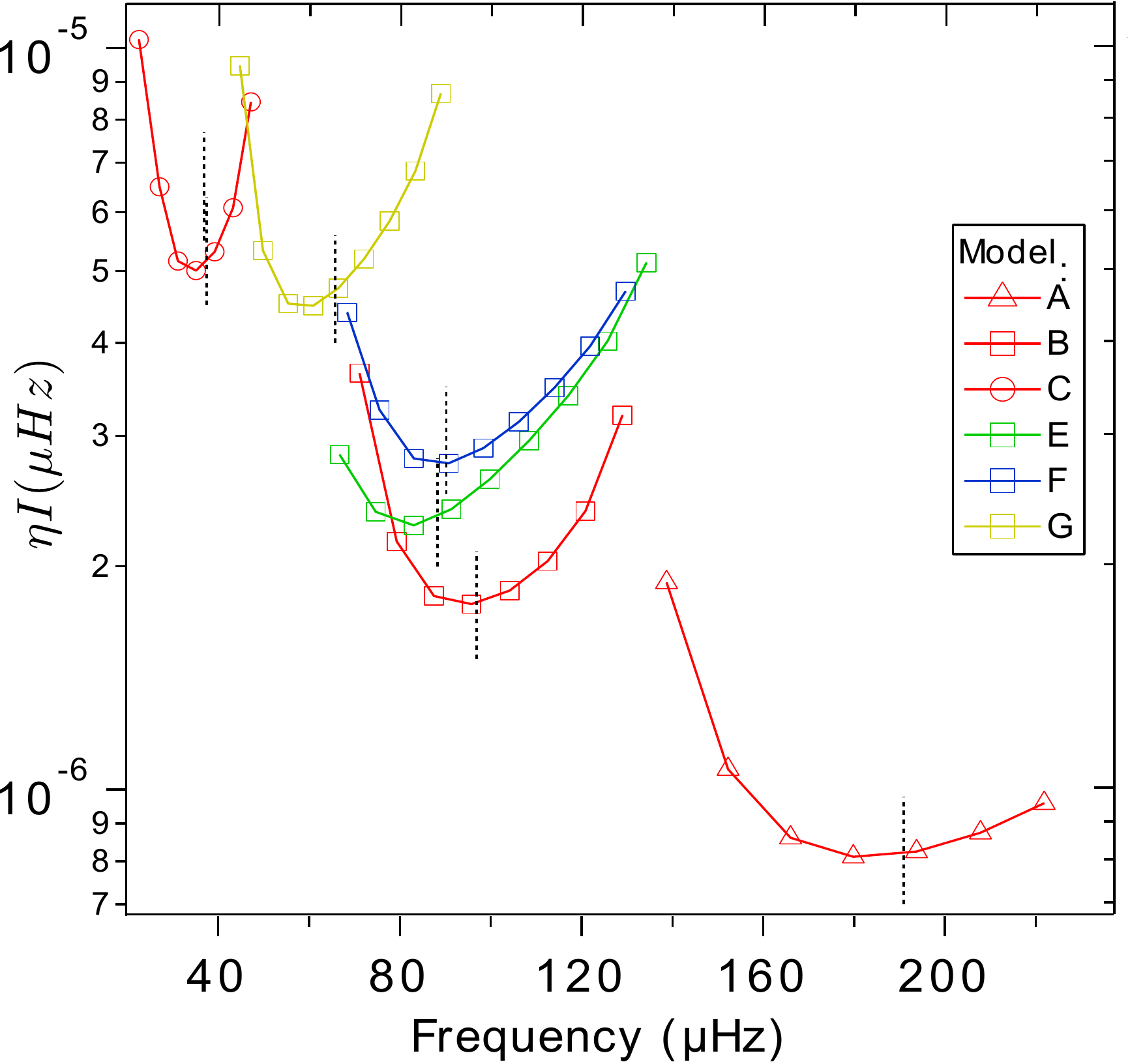}
      \caption{Product of the damping rates ($\eta$ in $\mu Hz$) by the dimensionless mode inertia ($I$) for radial modes for all our RGB models. The minimums of the curves correspond to the frequency of maximum height found in theoretical power spectra.  The dashed lines are the $\numax$ deriving from scaling relations.}
         \label{Fig:beta_all}
   \end{figure}

The main source of uncertainty in the TDC treatment comes from the closure term of the perturbed energy equation. This uncertainty appears in the form of a complex parameter $\beta$ \citep[Eq. 2 and Eq. 33 of][]{Grigahcene05}. \citet{Belkacem12} have shown that this parameter can be adjusted to obtain a plateau of the damping rates at the frequency $\numax$ predicted by the scaling relations  \citep[$\numax \propto g/T\ind{eff}^{1/2}$, first conjectured by][]{Brown}, which is similar to have a minimum in the product of the inertia and the damping rate. The existence of this plateau is well known in the solar case and is at the origin of the maximum observed in power spectra.
We thus adjust this parameter following the procedure of \citet{Belkacem12}, while paying particular attention to avoid non-physical spurious oscillations \citep{Grigahcene05}. 
In this procedure we assume that the canonical scaling relation for $\numax$ is valid. However, we know this relation is incomplete because, for example, the dependence to the Mach number is missing. Nevertheless, for red-giant stars, the dependence of $\numax$ to the surface gravity dominates, making the variation of the Mach number disappear during the evolution on the red-giant branch \citep[see][]{Belkacem_eta}.
In Fig \ref{Fig:beta_all} , using a unique value of $\beta$  ($\beta_{RGB} = -0.106 - 0.945 i$), we see that we can reproduce a minimum of $\eta I$ around the frequency $\numax$ predicted by the scaling relation for all our RGB models. For the helium burning model, we have to take another value of $\beta$ ($\beta_{RC} =  -0.130 - 0.950 i$).
This can seem to amount to only a small difference, but our predictions are very sensitive to this parameter.
We discuss the effect of this parameter on the lifetimes of the modes in Sect. \ref{appendix_tau}.
 We emphasize that such an approach makes our predictions more accurate than in \citet{Dupret09}, where the bell shape of the heights and the $\numax$ scaling relation were not reproduced.

A numerical difficulty occurring when computing red giant oscillation spectra comes from the existence of discontinuities of chemical composition and density in the equilibrium models. These discontinuities come from the evolution of the convective zones in the star and play an important role in mixed-mode trapping. The amplitudes of eigenfunctions at each side of the discontinuities strongly change from one mode to another. Therefore, we have adjusted the oscillation code to ensure the continuity of the Lagrangian perturbations of  pressure,  gravitational potential, and its gradient \citep[see also][]{Reese13}. Special care was also given to the computation of the Brunt-V\"{a}is\"{a}l\"{a} frequency as it significantly affects our results.

Another significant difficulty arises from the high density of non-radial modes over a large separation. It leads to having modes with very close angular frequencies (real part of modes eigenvalues) but with very different damping rates (imaginary part of the eigenvalues).  
The algorithm solving the non-adiabatic equations searches the eigenvalues by the inverse iteration method. Thus, it converges towards the closest eigenvalue to the initial guess in the complex plane. This eigenvalue is not necessarily the one corresponding to the frequency and trapping of the initial adiabatic mode. Using only the adiabatic frequencies as an initial guess for the real part of the eigenvalues,  the convergence to the correct mode (i.e the one with the frequency and trapping corresponding to the adiabatic case) of the algorithm is not easily ensured. Initial adiabatic frequencies of different modes could lead to the same eigenvalue in the non-adiabatic algorithm. As a remedy, we have to find an initial guess of the imaginary part for the frequency to be sure to obtain all the modes with different trapping. To do this, we use the inertia ratio between radial and non-radial modes derived from previous adiabatic calculations. We scale the initial guess for the imaginary part of the eigenvalues to the damping rates of radial modes with this inertia ratio.

\subsection{Stochastic excitation model}
\label{sect:stoc}
To compute mode heights, one also needs to compute mode driving. For this purpose, we consider the stochastic excitation model of \cite{SG01} \citep[see also][]{Belkacem06a,Belkacem06b,Samadi11} and consider the turbulent Reynolds stresses (hereafter $P_R$) as the dominant driving source. We do not take into account the entropy contribution (thermal source of driving) for which a severe deficiency in the modelling with non-adiabatic eigenfunctions appears \citep[as discussed in][]{Samadi13}.

We use solar parameters for the description of the turbulence in the upper convective layers \citep[constrained with a 3D numerical simulation by][]{Samadi03} with an extended Kolmogorov spectrum (EKS) for the $k$-dependency of the kinetic energy spectrum ($k$ is the wavenumber in the Fourrier space of turbulence) and a Lorentzian profile, with a high frequency cut-off to take into account the sweeping phenomena, for the eddy time-correlation function \citep{Belkacem10}. For the injection length-scale, we assume that it scales as the pressure scale-height at the photosphere \citep{Samadi08}.

With the power provided by the Reynold stresses and the damping rates from non-adiabatic computations we then compute the amplitude velocity ($V$) of the mode, using 
\begin{equation}
\label{eq:V2}
V^2= \frac{P_R}{2\eta M I}
\end{equation}
Note that one can convert the amplitude velocity into bolometric intensity following \citet{Samadi13}. A bolometric conversion taking into account the instrumental response was originally proposed by \citet{Michel09}. However, this approach is made in the adiabatic hypothesis. To make direct comparisons with observations, the visibilities of the mode should also be accounted for.
We give in Table \ref{tab:fT} the conversion factor to obtain the height for the radial mode at $\numax$ in ppm$^2$/$\mu$Hz. Neglecting here the non-adiabatic phase-lag, it is obtained using the relation 
\begin{equation}
\frac{\delta L}{L}=\frac{4 f_T - 2}{2\pi\nu R} V = C_f V
\end{equation}
with $f_T = \mid \delta T\ind{eff}/T\ind{eff} \mid / \mid \xi_r /R \mid$. We decide to present our results in radial velocity because the conversion to bolometric intensity introduces additional uncertainties.

\begin{table}[!h]
  \begin{center}
 \caption{Conversion factor $C_f^2$ from radial velocities to intensity variations for heights around $\numax$}
  \begin{tabular}{cccc}\hline 
{Model} & {$\numax [\mu Hz] $}&{$C_f^2$} & {H [$ppm^2/\mu Hz$] } \\  \hline
   A & 190 &$3.6\times 10^3$ &$9\times 10^3$  \\
   B & 97 &$2.9\times 10^3$ &$2\times 10^4$ \\
   C & 37 &$2.4\times 10^3$ &$8\times 10^4$ \\  \hline
   D & 37 &$3.2\times 10^3$ &$4\times 10^5$ \\ \hline
   E & 88 & $8.8\times 10^3$ &$4\times 10^4$ \\ 
   F & 90 & $12.7\times 10^3$& $6\times 10^4$\\ 
   G &  66 &$39.1\times 10^3$ & $3\times 10^5$ \\ \hline
  \end{tabular}
\label{tab:fT}
 \end{center}
\end{table}

To compute the height $H$ of a mode, we have to distinguish between resolved and unresolved modes. 
We assume that modes are resolved when their lifetimes ($\tau$) are small compared to the duration of observation ($\Tobs$), i.e.  $\tau \ll \Tobs/2$. In this case, we use for the height of the modes in the power spectra \citep[e.g.][]{Lochard05}
\begin{equation}
\label{eq:Hr}
H = V^2(R)\tau ,
\end{equation}
and for unresolved modes  $\tau \gg \Tobs/2$ 
\begin{equation}
\label{eq:Hur}
H_\infty = V^2(R)\Tobs/2 ,
\end{equation}
where $V(R)$ is the amplitude of the oscillation where it is measured (assumed here at the optical depth $\tau_R=0.1$), not including the disk integration factor. These formulae are strictly correct in the limit cases, when the lifetimes are much greater or lower than  the duration of observations. Both formulae give the same value for the height in $\tau = \Tobs/2$ but using Eq. \ref{eq:Hr} for $\tau<T_{\rm obs}/2$ and Eq. \ref{eq:Hur} for $\tau > T_{\rm obs}/2$, as done in \citet{Dupret09}, gives a derivative discontinuity in $T_{\rm obs}/2$. 

Note that the relation between amplitudes and heights depends on whether we deal with a two-sided  or a single-sided power spectrum. In the first case, the normalisation is such that the integral of the spectrum from $-\infty$ to $+\infty$ gives the total energy, so we have : 
\begin{equation}
V=\sqrt{\pi \Gamma H}
\end{equation}
In the opposite case, with a single-sided power spectrum we have
\begin{equation}
V=\sqrt{\pi \Gamma H/2}
\end{equation}
Here we consider a two-sided power spectrum, which is a different convention than the one used in \citet{Chaplin09}. Based on the formulation proposed by \citet{Fletcher} \citep[see also][]{Chaplin09} we have : 
\begin{equation}
\label{eq:H2}
H = \frac{V^2(R) \Tobs}{\Tobs/\tau +2},
\end{equation}
which tends to the same value of Eq. \ref{eq:Hr} an \ref{eq:Hur}  when   $\tau \ll \Tobs/2$ and $\tau \gg \Tobs/2$ respectively , and interpolate smoothly the heights between these two extreme cases. All our power spectra are computed using this latest description of the height.

With Eq. \ref{eq:H2} the height of the mode is only half its maximal height if $\tau = \Tobs/2$. More observational times  is required to fully resolve a mode. Indeed, in an observed power spectrum it is necesseray to have much more than two points within a linewidth to resolve the mode.

In this work we choose to take $\Tobs$ = 360 days. Despite the fact that we now have longer observational time with \emph{Kepler}, we picked up this value to discuss the different patterns that can occur in a power spectra and to have a clear distinction between resolved and unresolved modes in all our models. We will discuss the effect of increasing $\Tobs$ for each model in Sect \ref{sect:results}.

\section{Results}
\label{sect:results}
\subsection{Non-adiabatic effects on power spectra}
\label{sect:nadeffect}
The shape of the power spectra is mainly determined by two contributions: the modulation of inertia through mode trapping and the radiative damping. 
To discuss this shape, we will describe the behaviour of the ratio between the height of a g-type mode ($H_g$) and of a p-type mode ($H_p$). From Eqs. \ref{eq:V2}  and \ref{eq:H2} we have 
\begin{equation}
\label{HgHp2}
\left(\frac{H_g}{H_p}\right) =  \frac{(P_R I)_g}{(P_R I)_p}\frac{\eta_p I^2_p}{\eta_g I^2_g} \frac{f_g}{f_p} ,
\end{equation}
where $f_{g,p} =(\Tobs \eta_{g,p} +2)^{-1} $.
In the following, we derive this height ratio in the two asymptotic cases to discuss the main physical properties of the modes that can impact this ratio.

The following simple formulae illustrate this and can help with the interpretation of our results.
Assuming that the modes are resolved, $f_g/f_p$ in Eq. \ref{HgHp2} tends to $\eta_p/\eta_g$ and the height ratio is given by
\begin{equation}
\label{HgHg_r}
\left(\frac{H_g}{H_p}\right)_{res} =  \frac{(P_R I)_g}{(P_R I)_p}\left(\frac{(\eta I)_p}{(\eta I)_g}\right)^2 ,
\end{equation}
where $P_R I$ does not depend on the trapping  \citep[see][]{SG01} because the stochastic excitation is only efficient close to the surface (so $(P_r I)_g \simeq (P_r I)_p$). Taking into account the equation of the damping rate Eq. (\ref{damping}) and the decomposition of the work integral into the contribution of the core($W_c$) and the envelope ($W_e$) Eq. (\ref{eq:work}), we can rewrite the height ratio as
\begin{equation}
\label{eq:HgHP_r2}
\left(\frac{H_g}{H_p}\right)_{res} \simeq \left(\frac{(\eta I)_p}{(\eta I)_g}\right)^2 \simeq \left( \frac{(\int{dW})_p}{(\int{dW})_g} \right)^2    \simeq     \left(1+\left[\frac{ W_c }{W_e}\right]_g\right)^{-2}  , 
\end{equation}
where we use $\left(W_{e} \right)_g \simeq  \left(W_e\right)_p$  since the eigenfunctions of p-type and g-type modes are very close in the envelope. We also neglect the core contribution in the work integral of a p-type mode. We note from this formula that, when the radiative damping of g-type modes is negligible by comparison with the convective damping, their height are the same as the height of p-type modes if they are resolved; increasing the radiative damping clearly decreases the height ratio.

If we assume now that the p-type mode is resolved and the g-type mode is not (which is often the case in observed power spectra), the situation is different and $f_g/f_p$ in Eq. \ref{HgHp2} tends to $\eta_p \Tobs/2$ so
\begin{equation}
\label{HgHp_ur}
\left(\frac{H_g^{unres}}{H_p}\right) =  \frac{(P_r I)_g}{(P_R I)_p}\frac{(\eta_p I_p)^2}{\eta_g I_g^2}\frac{\Tobs}{2} = \left(\frac{(\eta I)_p}{(\eta I)_g}\right)^2\eta_g\frac{\Tobs}{2} ,
\end{equation}
following the same development as in Eq. (\ref{eq:HgHP_r2}) we find

\begin{equation}
\label{HgHp_ur2}
\left(\frac{H_g^{unres}}{H_p}\right)\simeq \left(\frac{H_g}{H_p}\right)_{res}  \eta_p \frac{I_p}{I_g}\left( 1+\frac{\int_{core}{dW_g}}{\int_{env}{dW_g}} \right)\frac{\Tobs}{2} .
\end{equation}
We see from this formula that the height ratio of unresolved modes depends on both their inertia and radiative damping.

We can also rewrite the resolution criteria ($\Tobs/2 >\tau_g$) where $\tau_g = 1/\eta_g$ represents the lifetime of the mixed-mode with the maximum inertia, so that all modes are resolved.
\begin{equation}
\label{eq:Tobsdw}
\frac{\Tobs}{2} \eta_p \gg \frac{I_g}{I_p}\left( 1+\frac{\int_{core}{dW_g}}{\int_{env}{dW_g}} \right)^{-1} 
\end{equation}

We see in Eq. (\ref{eq:HgHP_r2}), (\ref{HgHp_ur2}) and (\ref{eq:Tobsdw}) that there are clearly two contributions, the inertia and the non-adiabatic effects, that determine the shape of the power spectra.
How the inertia depends on the models is detailed in several other studies  \citep[e.g.][]{Montalban13} and can be understood through simple asymptotic derivations  \citep[see][and Appendix \ref{Appendix_inertia}]{Goupil13}.
The work due to non-adiabatic effects for p-type mode can be estimated with scaling relations \citep{Belkacem_eta}. The only remaining unknown is the ratio of the work integrals. We will discuss its evolution along with the evolution of the star in the next section.

\subsection{Power spectrum evolution of a $1.5 M_\odot$ star}

\label{sect:evol}
In this section, we will discuss the changes of the power spectrum along with the stellar evolution.  We will focus on the changes  of the damping rates and on the height of mixed-modes, as well as the effect of increasing the duration of observation.
We consider the $1.5 M_\odot$ models as described in  Sect.~\ref{sect:models}.

\begin{figure}[!h]
	\centering	
	\includegraphics[width=1.\linewidth]{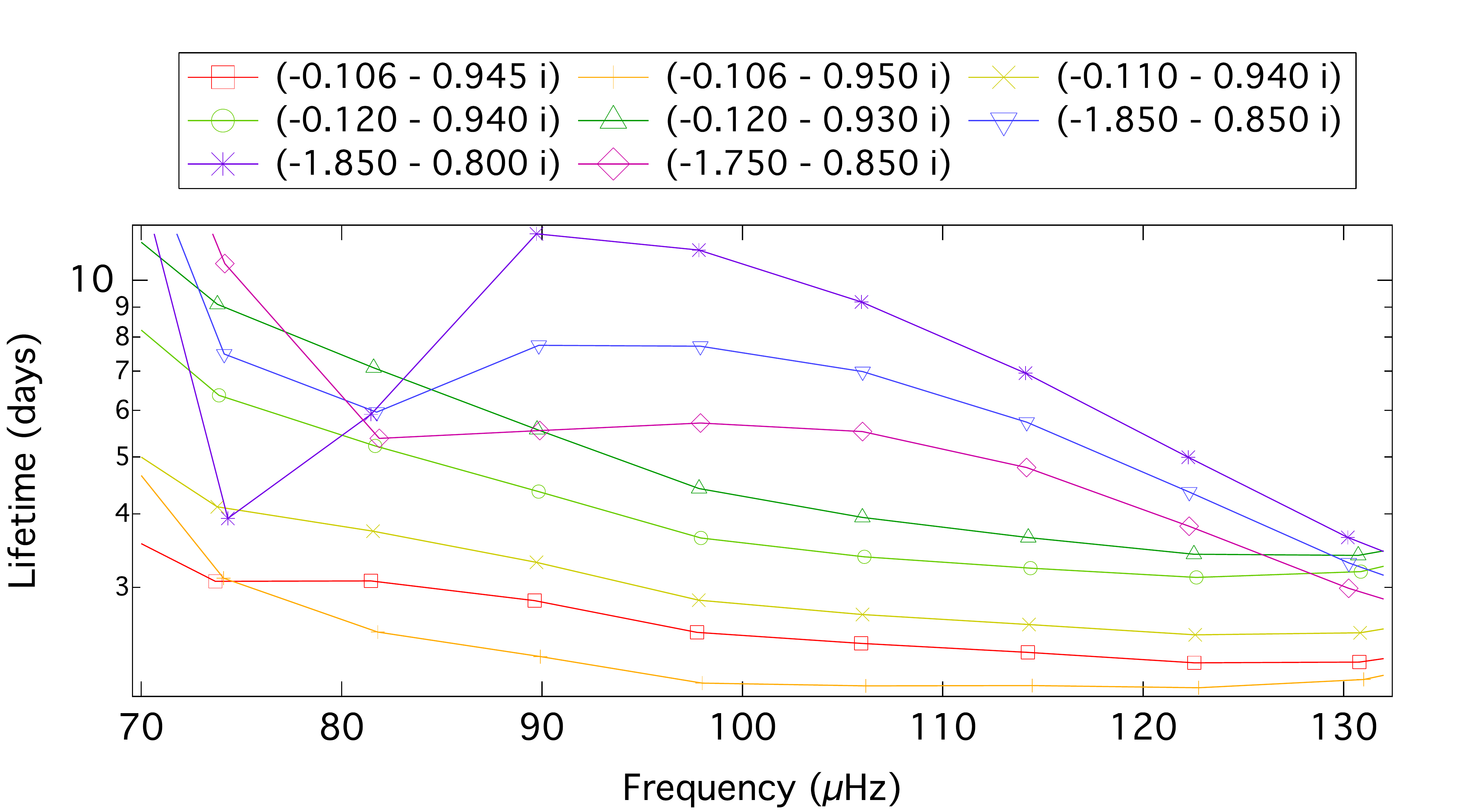}%
	\caption{Theoretical lifetimes of radial modes of model B for different values of $\beta$. The lifetimes corresponding to $\beta$ used in this paper are in red squares.}
        \label{tau-beta}
\end{figure}

\label{appendix_tau}
The theoretical lifetimes of the p-type modes can be strongly affected by the choice of the complex parameter $\beta$ in the time-dependant convection treatment.
We present in Fig.\ref{tau-beta} the lifetimes of the radial modes for the model B for different values of the $\beta$ parameter. All the values of $\beta$ presented give a minimum of $\eta I$ around the frequency $\numax$ predicted by the scaling relation.

 \begin{figure*}[ht]
       \includegraphics[width=1.\linewidth]{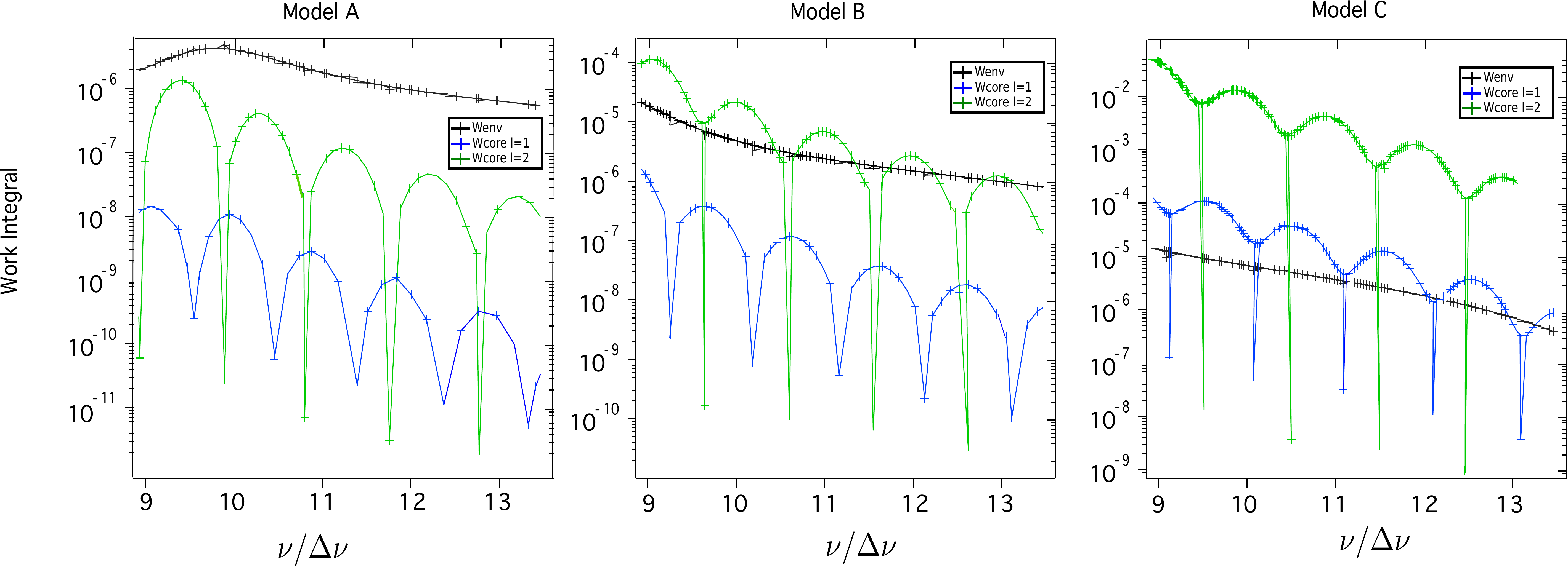}
      \caption{ Core and envelope contributions to the work integral for the three RGB models of $1.5 M_\odot$. The work is normalised by $GM^2/R$. Unlike the envelope contribution, the core contribution depends on the trapping and on the angular degree of the mode.}
         \label{fig:work}
  \end{figure*}

We first computed the lifetimes of radial modes for many values of $\beta$ throughout
the complex plane. Based on this investigation, we located a value of $\beta$  which gives
a minimum of $\eta I$ at $\numax$ for all our RGB models. We first present the results
for this $\beta$.
The power spectra for other values of $\beta$ are presented at the end of this section.

\begin{figure}[h]
     \includegraphics[width=1.\linewidth]{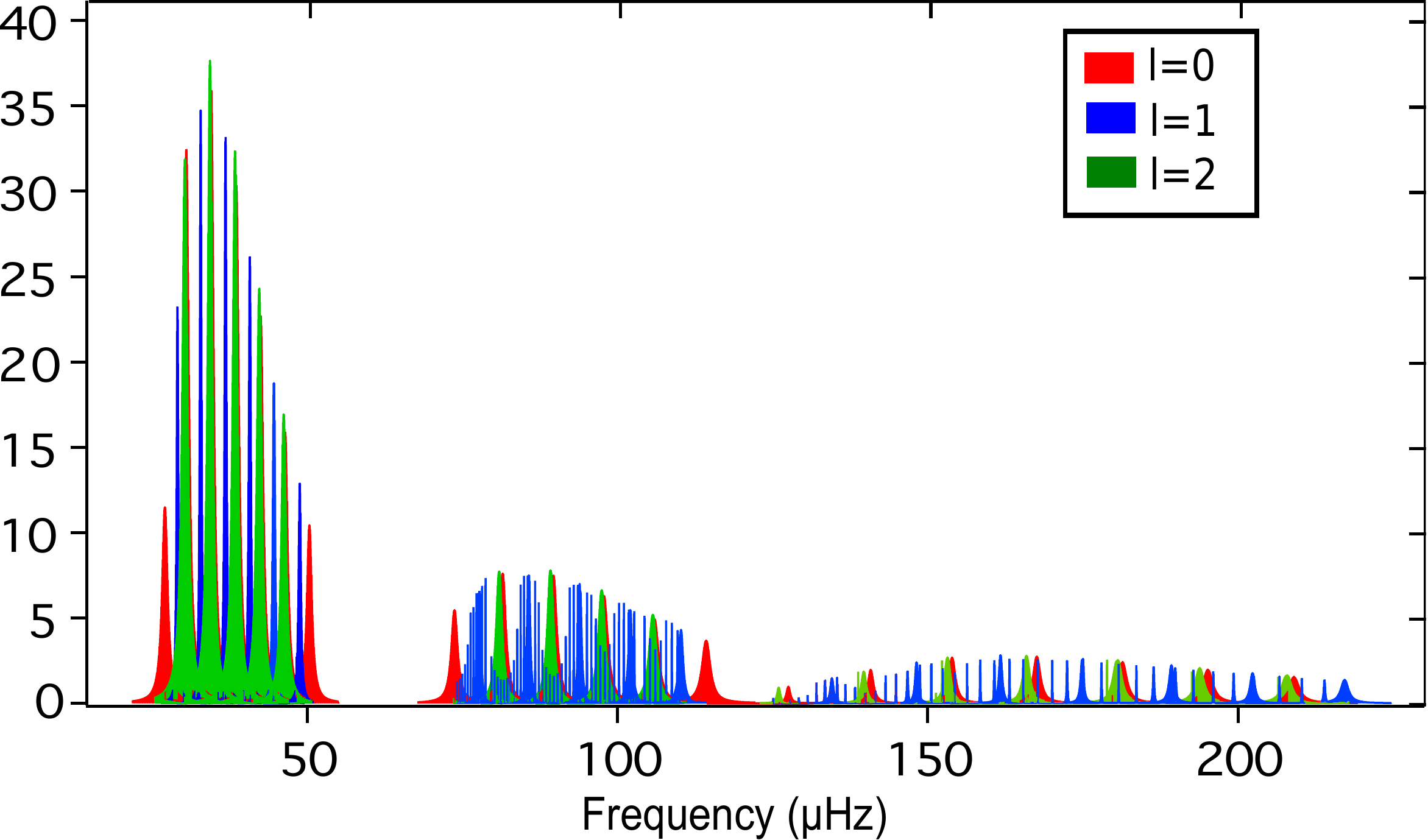}
      \caption{ Power spectra of the three RGB models (A,B and C, from right to left) of $1.5 M_\odot$, summerazing the gobal evolutionnary tendencies on power spectra: When the star evolves on the RGB, the solar-like oscillation range goes to lower frequency, the large separation decreases, the period spacing decreases and the height of the modes in the power spectrum increases.}
         \label{fig:allpow}
  \end{figure}

The work integrals of models A to C are presented in Fig.~\ref{fig:work}, while the corresponding mode lifetimes and theoretical power spectra are displayed in Fig.~\ref{fig:M15-all}.  As an overall tendency we find in  Fig. \ref{fig:allpow} the well known global changes of the power spectra with the evolution of the star on the RGB (the frequency range of solar-like oscillations goes to lower frequency and, the large frequency and the period spacing decrease).  In particular, we find an increase of the heights in the power spectra during the evolution of the star qualitatively compatible with previous theoretical computations \citep{Samadi2012} and with observations \citep[][]{Mosser12}.

For all our models, we show the lifetimes (Fig.~\ref{fig:M15-all} and Fig.~\ref{fig:M-all} left panel) for the radial and non-radial modes around $\numax$ and the associated power spectra in Fig.~\ref{fig:M15-all} and Fig.~\ref{fig:M-all} (right panel). These synthetic power spectra take into account the resolution of the modes for the height of the peaks. 
 However, we do not take the noise background into account (which can limit  the detectability of mixed-modes). So, we consider that the detectability of mixed-modes can be derived from the appearance of peaks in our synthetic power spectra. 
For convenience, we model all peaks by Lorentzians, which give us a limit power spectrum.

\begin{figure*}[t]
	\centering	
	\includegraphics[width=1.\linewidth]{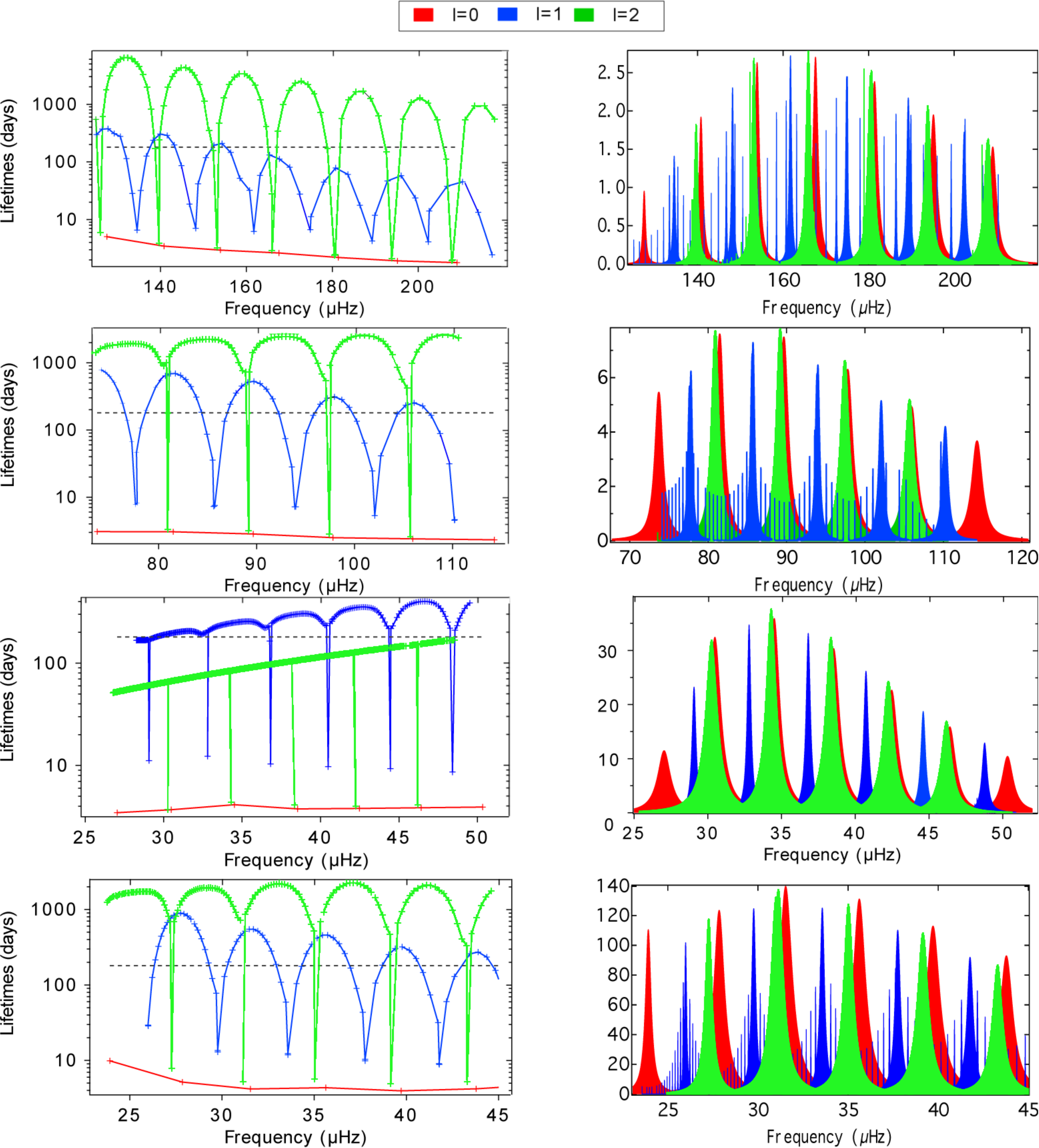}
	 \caption{\textbf{Left :} Lifetimes of $\ell=0$ (red), $\ell=1$ (blue) and $\ell=2$ (green) modes in models A, B, C and D (from top panel to bottom). The dashed line represente $\Tobs/2$.\textbf{Right :} Corresponding power spectra. The heights in power spectra are given in $(m/s)^2/\mu$Hz. For the sake of simplicity, all peaks are modelled by a lorentzian and we did not account for the different mode visibilities that depend on the angular degree. The resolution criteria (and so the time of observation) is taken into account for the height  of the modes. }
        \label{fig:M15-al-eps-converted-to.pdf}
\end{figure*}

\paragraph{Model A:}
At the bottom of the red-giant branch, the radiative contribution in the work integral (Eq.~\ref{eq:work}) is small for all modes by comparison with the convective one (Fig.~\ref{fig:work}, left panel). In addition, the convective work is a smooth function of the frequency and thus, as already pointed out in Sect.~\ref{sect:nonad}, is independent of the mode trapping. Hence, in the damping rate equation (see Eq.~\ref{damping}), only the inertia is responsible for the observed modulation of dipole and quadrupole modes lifetimes (Fig.~\ref{fig:M15-all}, first panel).

In this model, all dipole mixed modes are resolved (except for low frequencies) and have amplitudes high enough to be detected. Since the radiative damping is always negligible, their heights are close to those of the modes trapped in the envelope (p-dominated and radial modes, see Eq. \ref{eq:HgHP_r2}). Note that this spectrum is more regular than the one corresponding to model A in \citet{Dupret09}.
Some quadrupole mixed-modes, close to the p-dominated ones, are also visible in the synthetic power spectrum  (Fig \ref{fig:M15-all}, first panel). 

Increasing the time-duration would not change the dipole mode profiles. Indeed, since those modes are resolved, their heights no longer depend on the duration of the observations. Thus, at this early stage on the red-giant branch, we already find a clear structure in the power spectrum for dipole modes allowing us to derive a period spacing. Conversely, as the observation duration increases, the number of visible quadrupole modes increases too. With four years of observation, some quadrupole modes are resolved (more precisely quadrupole modes with lifetimes lower than 700 days) with heights comparable to the $p$-dominated modes. Moreover,  the height of some quadrupole unresolved modes increases so as to become visible in the synthetic power spectra. Finally, very long time of observation (typically about $27.5$ yrs) is required to have all quadrupole mode resolved, with all heights similar to the heights of the radial ones.

In this star, we also check the behaviour of the $\ell = 3$ modes (not represented on the figures). At this early stage on the RGB, they already undergo a strong radiative damping so that only the modes trapped in the envelope are visible in the power spectrum  (observations longer than hundred years would be required to see them). The increase of the radiative damping during the ascension of the red giant branch will  prevent even more to detect $\ell = 3$ g-types modes higher on the RGB. We thus predict that the detectable $\ell=3$ modes in red giants are all p-type modes.

\paragraph{Model B:}

Higher on the red-giant branch, the radiative contribution to the work integral is of the same order as the convective contribution for quadrupole modes (Fig.~\ref{fig:work}). This explains why the lifetimes for low frequency quadrupole modes level off (Fig.~\ref{fig:M15-all}, second panel). Moreover, the coupling between the two cavities decreases due to the contraction of the core and the expansion of the envelope. Indeed, when the star evolves, the number of mixed modes by large separation increase leading to an increase of the inertia ratio between a p-type and a g-type mode (see  Appendix \ref{Appendix_inertia}). Because of these two effects, $\ell=2$ mixed-modes are no longer visible in our synthetic power spectrum. By increasing the duration of observation above two times the lifetime of quadrupole modes (corresponding to approximately 10 years of observation), they would still not be detectable due to their significant radiative damping in the core.

\begin{figure}[!h]
	\centering	
	\includegraphics[width=1.\linewidth]{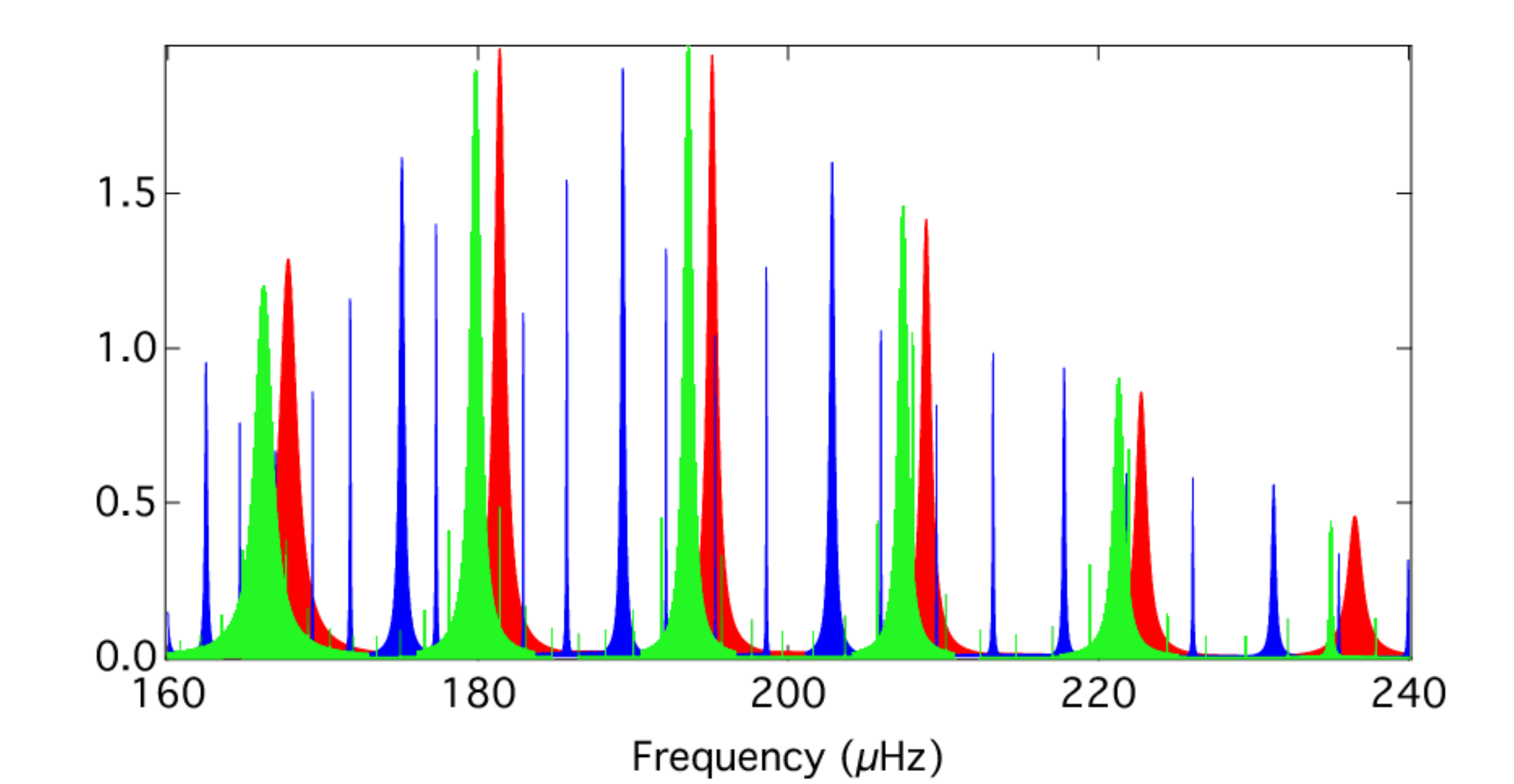}\\ %
	\includegraphics[width=1.\linewidth]{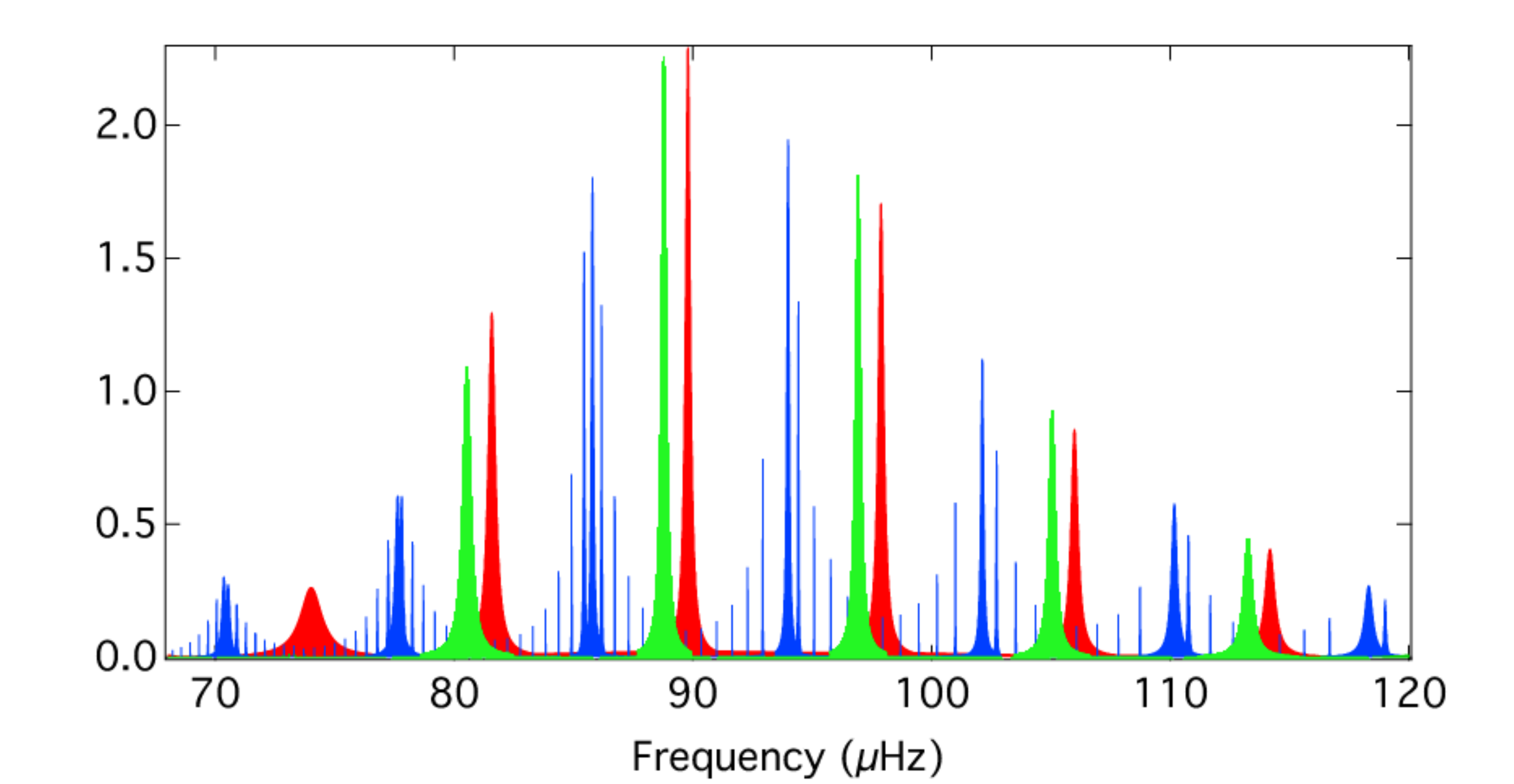}\\%
	\includegraphics[width=1.\linewidth]{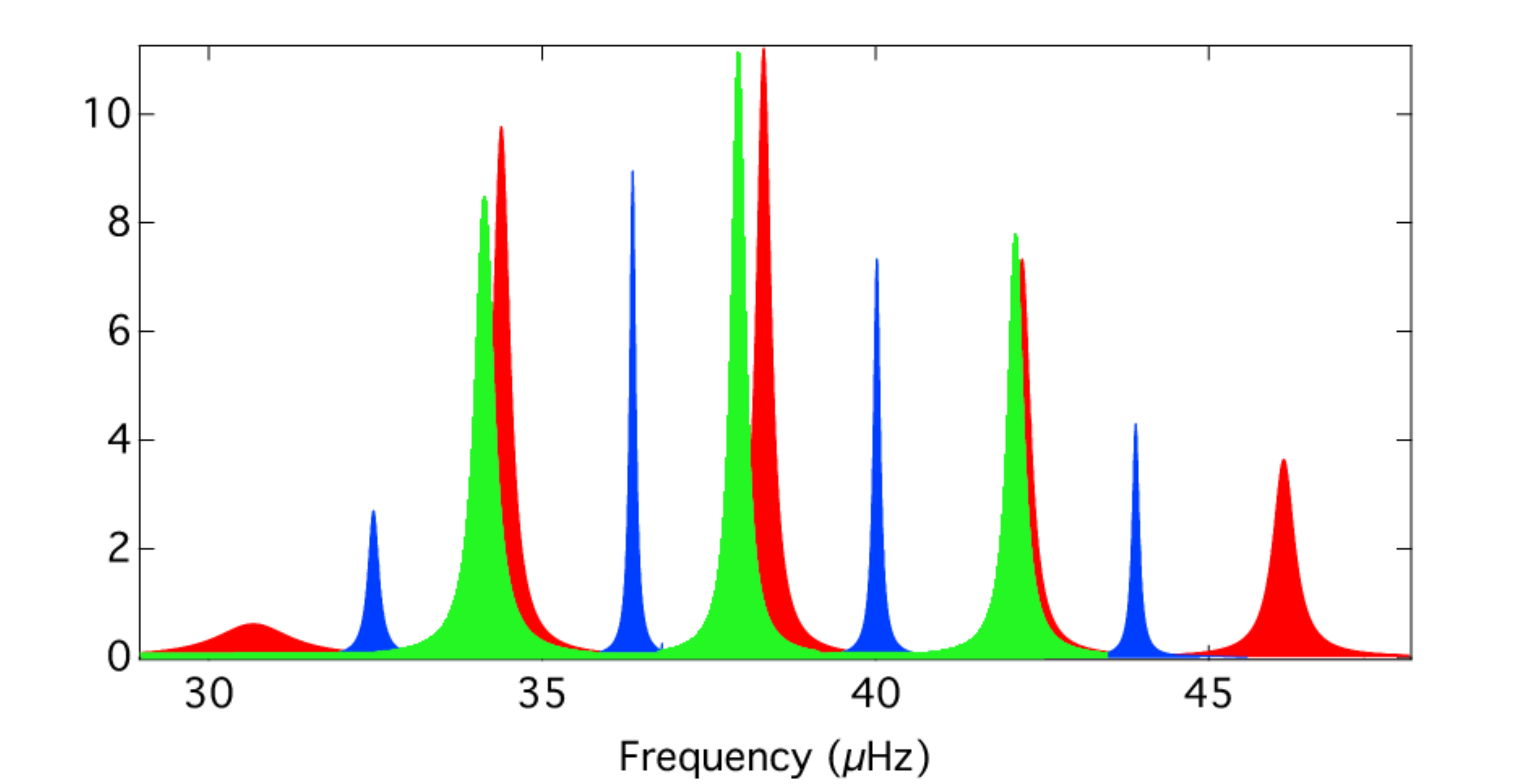}%
	\caption{From top to bottom: power spectra of models A, B and C with new values of the $\beta$ parameter allowing to have longer lifetimes for the p-type modes ($\beta_A = 1.700 -0.700 i $, $\beta_B = -1.940 -0.800$, $\beta_C  = -1.780 -0.920 i$). The detectability of the g-type modes is not significantly affected by the change of this parameter.}
        \label{pow-beta2}
\end{figure}

For dipole modes, the convective contribution is still the dominant part of the work integral so that their lifetimes are still clearly modulated by the inertia. Dipole modes strongly trapped in the core are not resolved and have smaller amplitudes. Moreover, as shown in Fig.~\ref{fig:M15-all} (second panel), their detection would be made difficult by the overlapping with radial modes and p-type quadrupole modes that exhibit  large linewidths. We detail in the end of this section the effect of the TDC parameter $\beta$ on the lifetimes of the p-type modes (see also Fig \ref{tau-beta}). Other values of this parameter could lead to higher lifetimes and hence, to narrower peaks (see Fig \ref{pow-beta2}).
Nevertheless, increasing the duration of observations will increase the heights of dipole modes. Taking 4 years of observation would allow us to have almost all $\ell=1$ modes resolved and in this case, their heights are very similar to the p-dominated non-radial modes.

\paragraph{Model C:}

For a more evolved model, the radiative damping continues to increase and the coupling between the two cavities becomes very small due to the expansion of the envelope and contraction of the core. This implies that the lifetimes of all modes,  except modes strongly trapped in the envelope, are dominated by the radiative damping (Fig.~\ref{fig:M15-all}, third panel). This damping is high enough to obtain lifetimes of g-dominated quadrupole mixed modes lower than the dipole ones. Consequently, only p-dominated modes are detectable (Fig.~\ref{fig:M15-all}, third panel). In this model, increasing even more the duration of the observation (even with $\Tobs > 2\tau$ for all modes) does not lead to detectable mixed modes, because of strong radiative damping (much more important than the convective one, Fig.~\ref{fig:work}).

\paragraph{Model D:}
 
Further on the evolution, after the helium-flash, the star begins to burn helium in its core. This model presents lifetimes similar to those of model B (Fig.~\ref{fig:M15-all}, fourth panel). After the helium-flash, the core has expanded and the envelope contracted leading to a decrease of the radiative damping of mixed modes and a stronger coupling between the p and g cavities. The appearance of a convective core also contributes to this decrease. The detectability of mixed modes (Fig.~\ref{fig:M15-all}, fourth panel) is very similar to the case of model B . After the He-flash, the radiative damping of the $\ell=3$ modes is still too high to see the g-type modes in our synthetic power spectrum.

\paragraph{Change of $\beta$ parameter:}
With the value of $\beta$ presented above (the same for all the RGB models), the resulting lifetimes  seem smaller than observations \cite[see e.g.][]{Baudin2011,Appourchaux2012}. We present in Fig \ref{pow-beta2} the power spectra obtained for other values of the $\beta$ parameter. In this case, we obtain higher lifetimes but at the price to change the value of $\beta$ between every model to be able to reproduce the scaling relation for $\numax$. This change of parameter does not affect much the general aspect of the power spectra and in particular the detectability of g-type modes.
Comparing the predicted lifetimes for stellar models fitting a selection of stars to observations would allow us to derive better constraint for the $\beta$ parameter.

\subsection{A proxy to the shape of power spectra}
\label{sect:discussion}

In the previous section, we have seen that when the stars climb the red giant branch, the mixed modes become more difficult to detect until their heights are too low to be visible in our synthetic power spectra (even if we consider  enough time of observation to resolve all modes). 

\begin{figure}[h!]
     \includegraphics[width=0.95\linewidth]{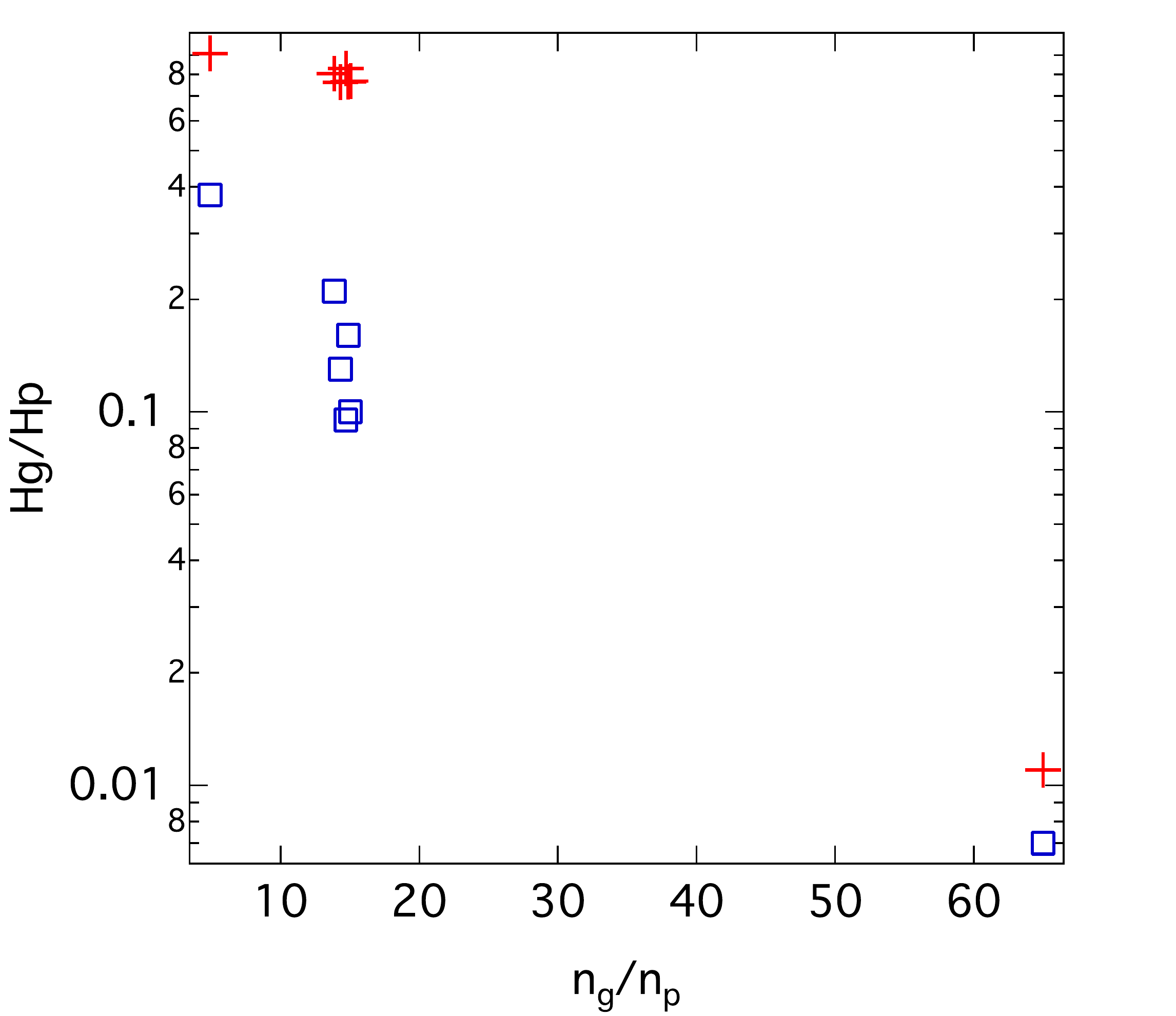}
      \caption{ Height ratio between a g-type (the $\ell=1$ with the highest inertia close to $\numax$) mode and a p-type mode around $\numax$ as a function of $n_g/n_p$ for all our models. Red crosses represents the ideal case where all the modes are resolved. Blue squares are for observation durations of one year.
 }
         \label{fig:HgHp}
  \end{figure}

\begin{figure*}[t]
	\centering	
	\includegraphics[width=1.\linewidth]{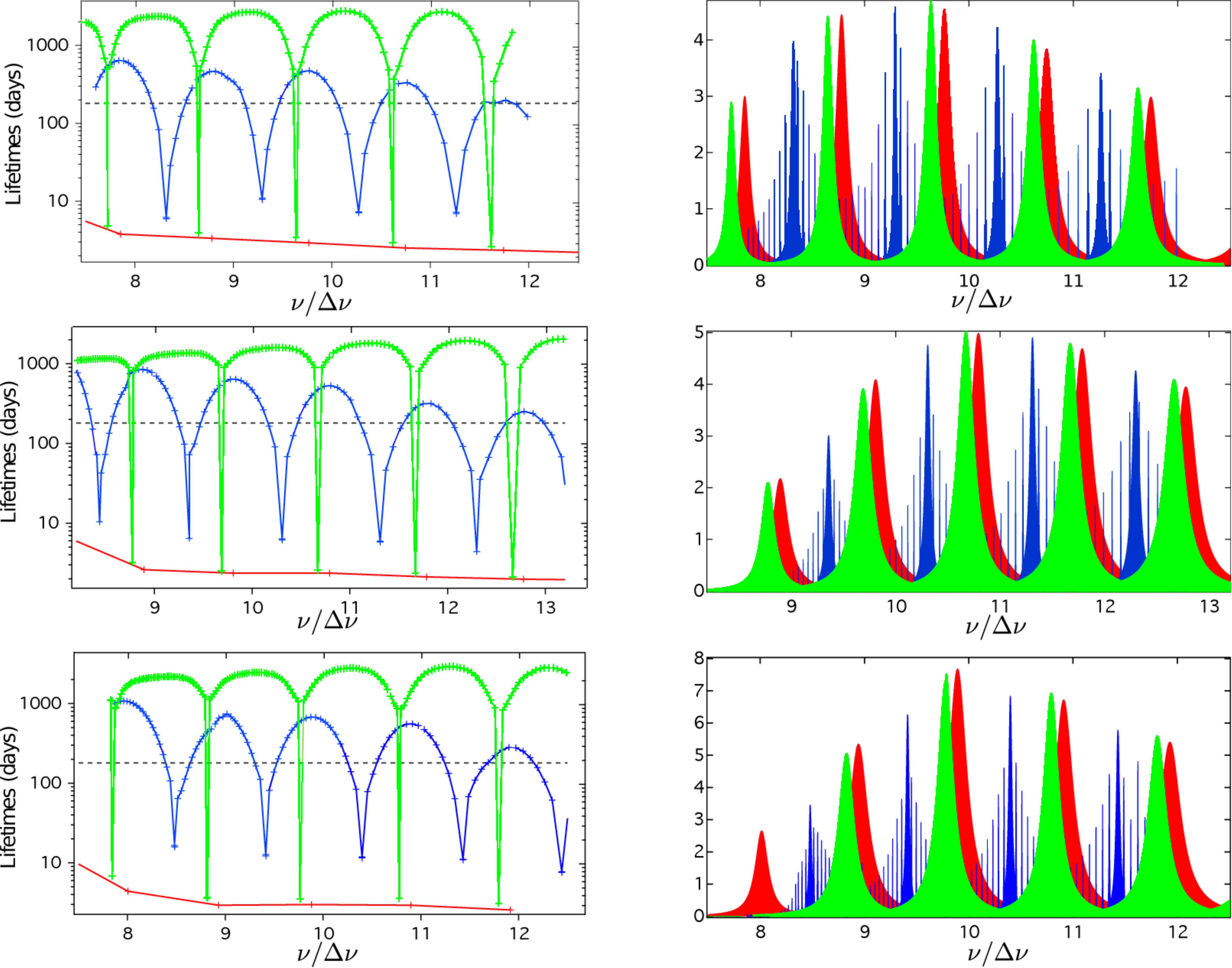}
	 \caption{\textbf{Left :} Lifetimes of $\ell=0$ (red), $\ell=1$ (blue) and l=2 (green) modes in models E,F and G (from top panel to bottom). The dashed line represente $\Tobs/2$.  \textbf{Right :} Corresponding power spectra. The heights in power spectra are given in $(m/s)^2/\mu$Hz. }
        \label{fig:M-all}
\end{figure*}

In the following lines, we note $n_g$ and $n_p$ the number of nodes of a dipole mode in the g- and p- cavities, respectively.
Our analysis of models with the same number of mixed modes over a large separation (Models E to G in Table~\ref{tab:models}), or equivalently a given ratio $n_g/n_p$, shows that they all exhibit the same behaviour for the lifetimes (Fig. \ref{fig:M-all}, left panel) . They also present very similar power spectra (Fig. \ref{fig:M-all}, right panel) with the same height ratios of mixed modes. 
Using the heights computed in the previous sections (i.e. from the calculations of the damping rates and of the Reynold stress) we show in Fig. \ref{fig:HgHp} a relation between the height ratio of g- and p-type modes around $\numax$ on the one side and $n_g/n_p$ on the other side, for fully resolved and partially resolved modes.
This relation is particullary marked in the case of fully resolved modes. More computations for a larger number of models  will be needed to derive a more precise relation for the relative heights of the modes.
 
 We notice that for unresolved modes, the inertia ratio can also be expressed as a function of $n_g/n_p$ (see Appendix \ref{Appendix_inertia}) so that this ratio is a good proxy for the shape of power spectra. In Fig.  \ref{fig:HgHp}  there is a higher dispersion between the models with the same $n_g/n_p$ with only one year for the duration of observations, because these modes are only partially resolved.
 We notice that a theoretical evaluation of this proxy is very easy through asymptotic relations and the scaling relation for $\numax$ : $n_g/n_p \simeq  \Delta \nu /\Delta\Pi \numax^2  $.
Taking then the background noise into account, it becomes possible to estimate the detectability of mixed modes along the red-giant branch.

\begin{figure}[h!]
     \includegraphics[width=1.\linewidth]{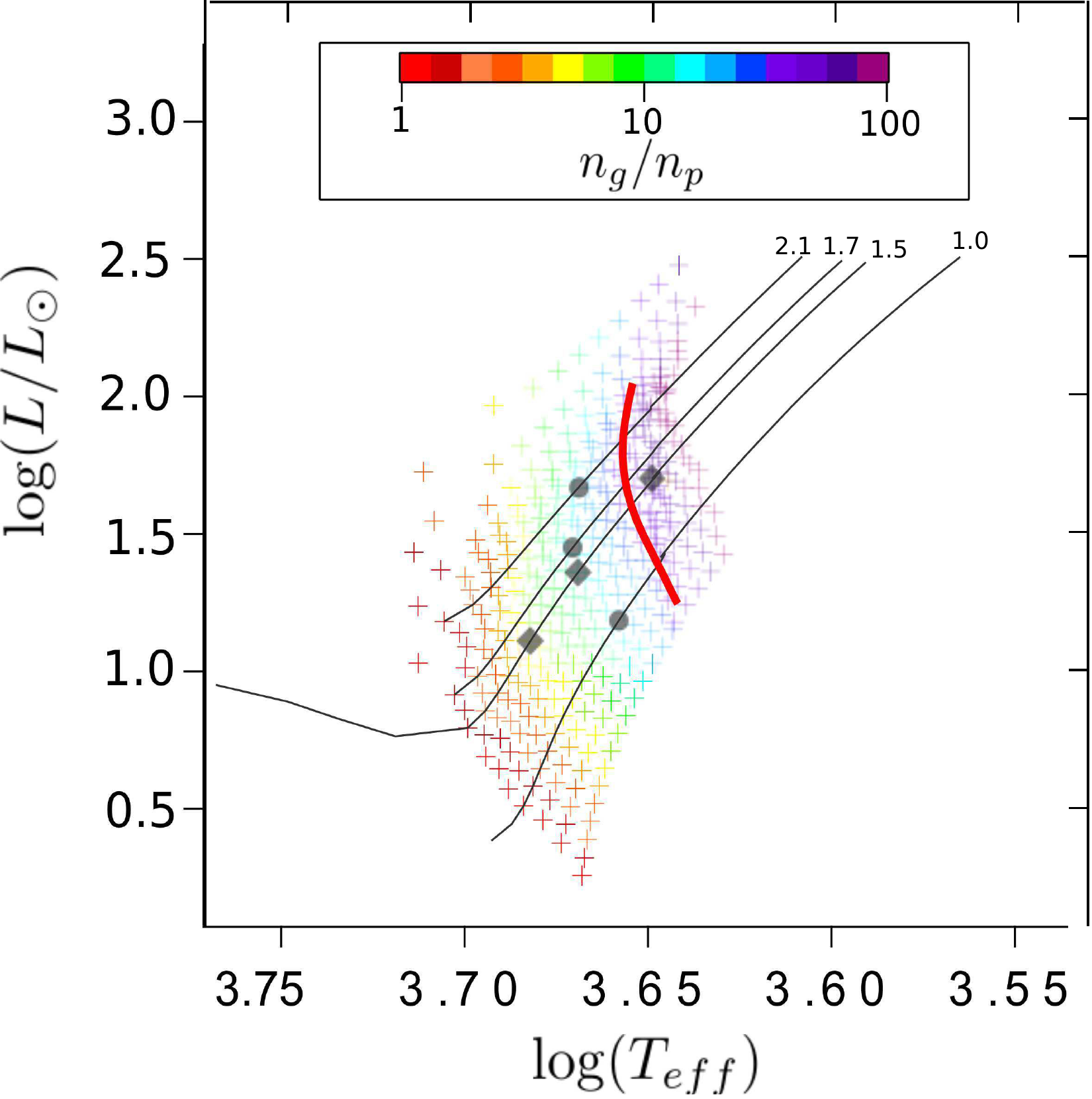}
      \caption{ Evolutionary tracks in the HR diagram of all our red-giant branch models. Numbers on the top of the tracks indicate the mass of the star (in $M_\odot$). The color scale indicates models with the same number of mixed modes by large separation. The red line represents the detectability limit we have found for the dipole modes (assuming that all modes are fully resolved).}
         \label{fig:HR_detect}
  \end{figure}

Using this relation, we present in Fig. \ref{fig:HR_detect} a red curve corresponding to $n_g/n_p \simeq 60$, which appears to be the level on the red-giant branch where we are no longer able to see any dipole mixed modes in the synthetic power spectra even by increasing the time of observation to more than 10 years (so that all modes are resolved).

\section{Conclusion}

We have determined lifetimes and heights of radial and non-radial mixed modes for several red-giant models of various evolutionary states and various masses.
The corresponding synthetic power spectra are in overall agreement with the observed bell shape. We follow the change of the frequency of maximal amplitude and reproduce the increase of the maximal height along with the evolution of the star.
For some of our models, we predict that long enough observations would increase the heights of mixed modes up to those of radial modes. But for models with larger density contrast between the core and the envelope, radiative damping becomes too strong and the coupling too weak to have detectable mixed modes in our synthetic power spectra.

In a $1.5 M_\odot$ star, we predict no detection of dipole mixed modes at $\numax \lesssim 50 \mu$Hz and $\Delta \nu \lesssim 4.9 \mu$Hz (corresponding to $n_g/n_p \gtrsim 60$) and no detection of quadrupole mixed modes around $\numax \lesssim 97 \mu$Hz and $\Delta \nu \lesssim 8.4 \mu$Hz (corresponding to $(n_g/n_p)_{\ell=2} \gtrsim 15$). We present in Appendix \ref{appendix_obs} a brief qualitative comparaison between the tendencies we found in our synthetic power spectra and Kepler spectra. Theoretical power spectra are in qualitative good agreement with observed ones. The general aspects of mixed-modes spectrum and it's evolution on the red-giant branch is well reproduced. The lifetimes of our p-type modes seems to be underestimated, but they strongly depend on the complex parameter $\beta$ (see Appendix \ref{appendix_tau} for more details). Quantitative comparisons between theory and observations will then require a much precise determination of this parameter. 

Computations of power spectra for other masses show that we have very similar lifetimes patterns and power spectra, and in particular the same height ratios for mixed modes, if we take models with the same number of mixed modes in a large separation. 
We can then rely on the number of mixed modes by large separation to predict the height ratio between a p-type and a g-type mixed mode for various stars.

More numerical computations varying not only the mass but also the other parameters of the equilibrium models (e.g. chemical composition, convection treatment, ...), will help to verify these results for a larger set of stars. 
Then we will have to compare our height ratios and their dependence with the number of mixed modes by large separation with a large set of observed power spectra \citep[see e.g.][]{Mosser12}.
In such a large set of observations, some power spectra show depressed mixed modes.
Up to now, from a theoretical point of view, we did not find any depressed mixed mode in all our red-giant branch models. 
The discovery of models for which the inertia of p-type dipole modes as well as their damping rates are close to the g-type modes ones could give us a better insight in the mechanism depressing these modes but we did not find such models.

Our results should also be tested by comparing the observed height and lifetimes with the theoretical ones for some individual stars. 
Such comparisons can now be done with the measurements of linewidhts and heights of the modes in solar-like stars \citep[see e.g.][]{Appourchaux2014}  in subgiants  \citep[see e.g.][]{Benomar2013} and in red giants \citep{Huber2010,Hekker2010,Corsaro2012}. These comparisons will help in particular to test our TDC treatment and to calibrate the $\beta$ parameter. This will be the subject of a futur work.

\begin{acknowledgements}
This work is financially support through a doctoral fellowship from the F.R.I.A.
\end{acknowledgements}


\appendix
\section{Inertia ratios}
\label{Appendix_inertia}
To have a better understanding of the inertia ratios between a p-type and a g-type mode, we follow the development of  \citet{Goupil13}  based on the asymptotic method developed by \citet{Shibahashi79}.

In the asymptotic regime and neglecting the size of the evanescent zone, \cite{Goupil13} show that the inertia in the envelope of the stars varies as $I_{env} \simeq (c^2/2\pi\nu) \tau_p$ and in the core as $I_{core} \simeq (a^2/2\pi \nu) \tau_g$, so we can express the total inertia as 
\begin{equation}
I = I_{core} + I_{env} \simeq \frac{c^2}{2\pi\nu}(\tau_p +\frac{a^2}{c^2}\tau_g)
\end{equation}
with 
\begin{eqnarray*}
&&\tau_p = \frac{1}{\pi\nu} \int_{env} k_r dr \simeq \frac{1}{\Delta\nu} \\
&&\tau_g = \frac{1}{\pi\nu} \int_{core} k_r dr \simeq \frac{1}{\nu^2\Delta\Pi}
\end{eqnarray*}
with $k_r$ the radial wavenumber and $\nu$ the frequency of the mode in $\mu$Hz. $c$ is a normalisation constant and $a$ is related to $c$ by \citep[eq. (16.49) and eq. (16.50) from][]{Unno} :
\begin{equation}
\frac{c}{a} = \frac{2 \cos\left(arc\cot(\frac{1}{4}\cot(\pi\nu \tau_p))\right)}{cos(\pi\nu\tau_p)}
\end{equation} 
$(c/a)^2$ is then a function of $\nu$ of period $1/\tau_p = \Delta\nu$ which varies between 4 (p-modes) and 1/4 (g-modes).
Finally when comparing the inertia of a mode trapped in the envelope ($I_p$) and of a mode trapped in the core ($I_g$), we have
\begin{equation}
\label{eq:IgIp}
\frac{I_g}{I_p} \simeq \frac{1+4 \frac{\tau_g}{\tau_p}}{1+\frac{1\tau_g}{4\tau_p}}
\end{equation}
This inertia ratio is then a function of the ratio $\tau_g/\tau_p = n_g / n_p$, the number of mixed-modes by large separation.

\section{Qualitative comparaison to Kepler spectra}
\label{appendix_obs}
\begin{figure*}[t]
	\centering	
	\includegraphics[width=1.\linewidth]{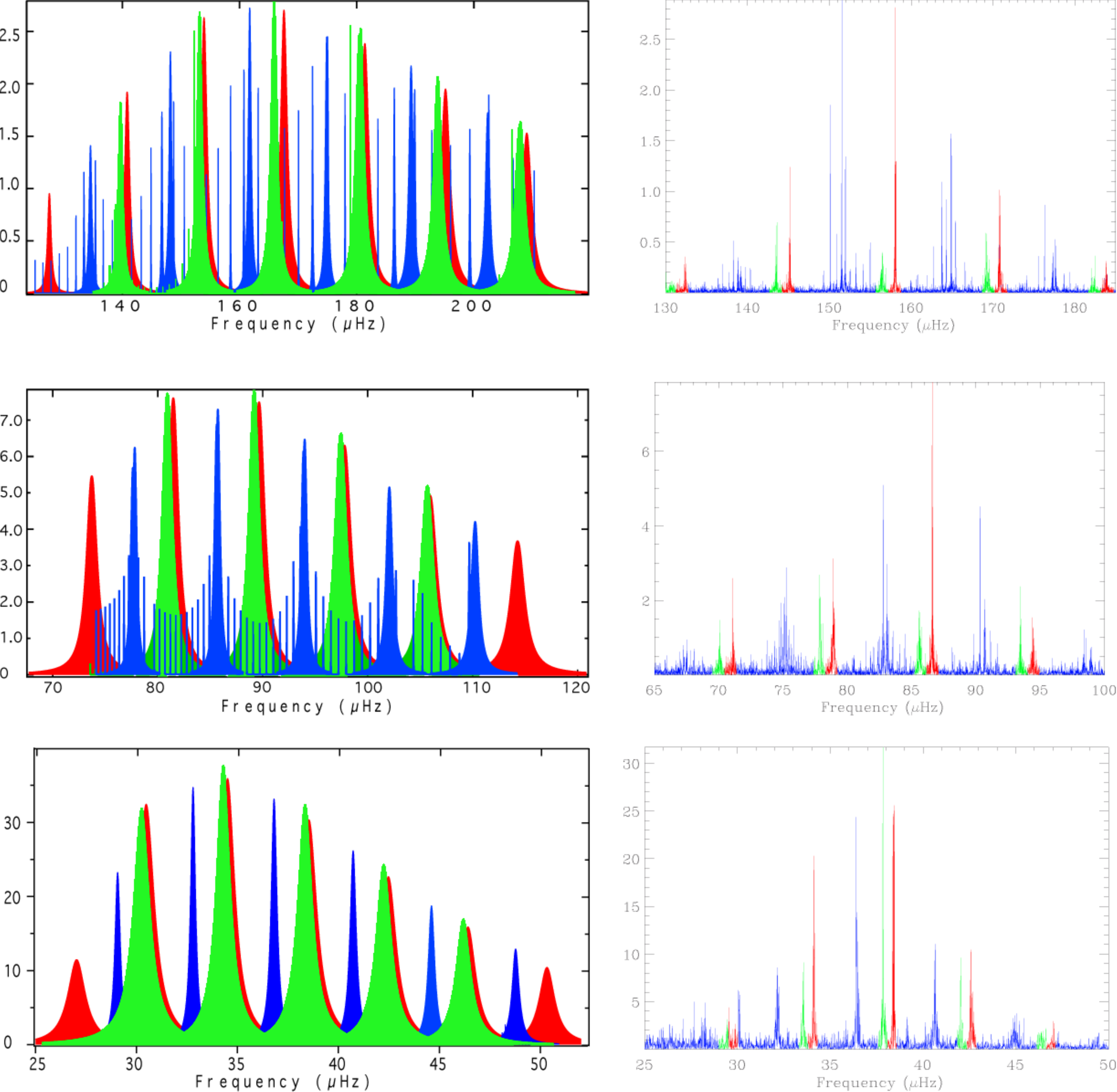}
	\caption{Theoretical and observed power spectra of Kepler stars with similar masses (from top to bottom  : 1.44, 1.48, 1.47 $M_\odot$), $\Delta\nu$ and $\numax$. The heights int theoretical power spectra are in $(m/s)^2/\mu Hz$. The heights for observed spectra are given in $ppm^2/\mu Hz$ divided by a factor $6000$ to have scales similar to the theoretical spectra.}
        \label{fig:pspecobs}
\end{figure*}

We present in Figure \ref{fig:pspecobs} some power spectra obtained with Kepler alongside with our $1.5 M_\odot$ RGB theoretical power spectra to show the main tendencies discussed in this paper.
Concerning the height ratios and the limit for the detectability of mixed-modes in our theoretical power specra, we found the same tendencies in the observed ones.
At the begging  of the red giant branch, dipole mixed-modes have heights comparable to p-type modes. Higher on the RGB, dipole mixed-modes are partially resolved and their heights  present a clear modulation compared to the heights of p-type modes. At the level of model C, only the p-type modes have significant heights.
The number of visible mixed modes is larger in the observed spectra, owing to the presence of rotational multiplets, but without consequence on their height and width.

\end{document}